\documentclass{article}\usepackage[]{graphicx}\usepackage[]{color}
\makeatletter
\def\maxwidth{ %
  \ifdim\Gin@nat@width>\linewidth
    \linewidth
  \else
    \Gin@nat@width
  \fi
}
\makeatother

\definecolor{fgcolor}{rgb}{0.345, 0.345, 0.345}

\usepackage{framed}
\makeatletter
 {\par\unskip\endMakeFramed%
 \at@end@of@kframe}
\makeatother

\definecolor{shadecolor}{rgb}{.97, .97, .97}
\definecolor{messagecolor}{rgb}{0, 0, 0}
\definecolor{warningcolor}{rgb}{1, 0, 1}
\definecolor{errorcolor}{rgb}{1, 0, 0}
\newenvironment{knitrout}{}{} 

\usepackage{alltt}
\usepackage{amsmath}
\usepackage{amssymb}
\usepackage{graphicx,psfrag,epsf}
\usepackage{enumerate}
\usepackage{natbib}
\usepackage{url} 
\usepackage{fullpage}
\usepackage{MnSymbol}
\usepackage{bm}
\usepackage{tikz}
\usepackage{picture}
\usepackage[linesnumbered,lined,commentsnumbered]{algorithm2e}
\usetikzlibrary{shapes}
\usetikzlibrary{plotmarks}
\usepackage{floatrow}
\usepackage{hyperref}
\newfloatcommand{capbtabbox}{table}[][\FBwidth]

\IfFileExists{upquote.sty}{\usepackage{upquote}}{}
\begin{document}


\title{Bayesian prediction for physical models with application to the optimization of the synthesis of pharmaceutical products using chemical kinetics} 


\author{Antony~M.~Overstall$^\dagger$\footnote{Corresponding author. Southampton Statistical Sciences Research Institute, University of Southampton, Southampton, SO17 1BJ, UK. Email: A.M.Overstall@southampton.ac.uk}, David~C.~Woods$^\dagger$ \& Kieran~J.~Martin$^\ddagger$ \\[1ex]
$^\dagger$University of Southampton, Southampton SO17 1BJ, UK \\[.5ex]
$^\ddagger$Roche, Welwyn Garden City AL7 1TW, UK}






\date{}

\maketitle

Quality control in industrial processes is increasingly making use of prior scientific knowledge, often encoded in physical models that require numerical approximation. Statistical prediction, and subsequent optimization, is key to ensuring the process output meets a specification target. However, the numerical expense of approximating the models poses computational challenges to the identification of combinations of the process factors where there is confidence in the quality of the response. Recent work in Bayesian computation and statistical approximation (emulation) of expensive computational models is exploited to develop a novel strategy for optimizing the posterior probability of a process meeting specification. The ensuing methodology is motivated by, and demonstrated on, a chemical synthesis process to manufacture a pharmaceutical product, within which an initial set of substances evolve according to chemical reactions, under certain process conditions, into a series of new substances. One of these substances is a target pharmaceutical product and two are unwanted by-products. The aim is to determine the combinations of process conditions and amounts of initial substances that maximize the probability of obtaining sufficient target pharmaceutical product whilst ensuring unwanted by-products do not exceed a given level. The relationship between the factors and amounts of substances of interest is theoretically described by the solution to a system of ordinary differential equations incorporating temperature dependence. Using data from a small experiment, it is shown how the methodology can approximate the multivariate posterior predictive distribution of the pharmaceutical target and by-products, and therefore identify suitable operating values. Materials to replicate the analysis can be found at \url{www.github.com/amo105/chemicalkinetics}.\\[1ex]




\section{Introduction}
\label{INTRO}

Pharmaceutical products are often manufactured by chemical synthesis, with an initial set of chemical substances evolving over time via a series of chemical reactions into a substance or substances of interest. One (or more) of these substances will be the target pharmaceutical product while the others will be unwanted or harmful \textit{by-product} substances. In addition to a time dependence, the chemical synthesis process may be hypothesised to depend on various controllable factors. The aim of the chemical engineers will be to manipulate the controllable factors such that specification limits on the final substances are satisfied. For example, the quantity of pharmaceutical product may be required to exceed some specified level whilst the amounts of unwanted by-products are correspondingly kept below a certain level.

It is common for a physical model, for example derived from scientific theory as the solution to a set of ordinary differential equations (ODEs), to be postulated to approximately describe the dependence of the synthesis process on the controllable factors (and time); see chapters in \citet{amEnde2011}. In addition, often an experiment can be performed within which the amounts of the substances of interest are measured for several different specifications of the controllable factors. These observed responses can then be used to inform the relationship between controllable factors and the synthesis process, e.g. through the estimation of unknown tuning parameters.

In this paper, we develop and apply methodology for the statistical modelling of such experiments, motivated by a pharmaceutical exemplar which measured three substances of interest: the pharmaceutical product and two unwanted by-products. The dynamics behind the synthesis process are approximately described by a system of non-linear ODEs, the solution to which forms a physical model for the amounts of the three substances at a given time. This model depends on a set of controllable factors (including time) and unknown parameters. A small experiment has been conducted where the amounts of the substances of interest have been observed for different controllable factors at certain time points. The aim is to use this observed data to estimate the unknown parameters and then to use these estimates to maximize the probability of the predicted amounts satisfying known specification limits as a function of the controllable factors and time. The approach used here will be Bayesian, giving a coherent method of propagating uncertainty on the unknown parameters (given the observed data) through to the predicted amounts. This approach is closely aligned with the concept of ``pharmaceutical design space''\footnote{\url{https://www.fda.gov/downloads/drugs/guidances/ucm073507.pdf}}, a set of combinations of values of the controllable factors that have been demonstrated to provide assurance of quality. See, for example, \citet{peterson2008}, \citet{peterson_yahyah2009} and \citet{Lebrun_et_al2013}  for related Bayesian approaches to the definition of a design space using linear statistical modelling methods. Beyond pharmaceutical design space, in the wider field of process optimization, see, for example, \citet{chiao2001}, \citet{peterson2004} and \citet{delcastillo2007}.

Despite the apparent simplicity and appealing nature of the Bayesian modelling approach, several computational challenges remain. Firstly, the solution to the system of ODEs is analytically intractable and so a computationally expensive numerical solution is required \citep[e.g.][]{I2009}. The computational expense of this approximation will impact the computing time required both for the estimation of the unknown parameters through the evaluation of their posterior distribution \textit{and} the predictions for new combinations of the controllable factors. To reduce computational expense, we make use of (multivariate) Gaussian process emulators to accelerate both the Markov chain Monte Carlo (MCMC) algorithm employed and the prediction of future responses. Secondly, some of the observations fall below a minimum threshold, $\chi$, under which the actual amount cannot be measured reliably, resulting in left-censored observations. Instead of assuming that the true observations are equal to zero or $\chi$ with certainty, we assume that the true observations are unknown, subject to being in the interval $(0,\chi)$, within a natural Bayesian formulation. Thirdly, we know that the physical model derived from the solution to the ODEs only provides an approximation to the true process. We account for any potential mismatch by introducing an explicit model discrepancy term \citep[c.f.][]{KOH2001}. Fourthly, the function giving the probability of satisfying the specification limits for given controllable factors will be noisy and computationally expensive to evaluate making optimization non-trivial. 

The contribution of this paper is the novel application of a combination of modern computational and statistical methodology to address these challenges, and the demonstration of this approach on an important practical problem.

The paper is organised as follows. In Section~\ref{BACK} we describe the motivating experiment, and the physical and statistical models for the observations in more detail. In Section~\ref{METH} we describe the methodology including construction of Gaussian process emulators. In Section~\ref{RES} we present the results of applying the methodology to the motivating experiment, before concluding with a discussion in Section~\ref{DISC}.

\section{Background}
\label{BACK}

\subsection{Chemical Reactions}
\label{ChemReact}
The motivating chemical synthesis process involves nine substances, labelled A to I. The substances of interest are E, F and H, where F constitutes the pharmaceutical product with E and H being unwanted by-product substances. Let $A(t)$ denote the amount (in mols) of substance A at time $t$ (in seconds), with similar notation for the remaining substances. At time $t=0$, initial amounts  of A, D and E, denoted by $A_0 = A(0)$, $D_0 = D(0)$ and $E_0 = E(0)$, respectively, are introduced into the laboratory apparatus at a specified temperature ($\lambda$ in Kelvins) and volume ($V$, in litres). Let $\mathbf{x} = (x_1,\ldots,x_5)=\left(A_0, D_0, E_0,\lambda, V\right)$ be an $F \times 1$ vector ($F=5$) and assume that $\mathbf{x} \in \mathcal{X} = \prod_{f=1}^F \left[x_{f,min},x_{f,max}\right]$. The upper and lower limits, $x_{f,min}$ and $x_{f,max}$, for each factor are given in Table~\ref{limits}.

\begin{table}[ht]
\caption{Lower and upper limits of the controllable factors in the motivating experiment.}
\label{limits}
\begin{center}
\begin{tabular}{llllll} \hline
Definition & Name & Lower limit, $x_{f,min}$ & Upper limit, $x_{f,max}$ & Units \\ \hline
Initial amount of A & $x_1 = A_0$ & 22.5 & 45 & mols \\
Initial amount of D & $x_2 = D_0$ & 91.4 & 91.59 & mols \\
Initial amount of E & $x_3 = E_0$ & 26.42 & 26.47 & mols \\
Temperature & $x_4 = \lambda$ & 298.15 & 313.15 & Kelvins (K)\\
Volume & $x_5 = V$ & 31.28 & 32.56 & litres (l) \\ \hline
\end{tabular}
\end{center}
\end{table}

The chemical synthesis process is governed by the following series of chemical reactions:
\begin{equation}
\left.
\begin{array}{lll}
\mathrm{A} & \stackrel{k_1}{\longrightarrow} & 2 \mathrm{B} + \mathrm{C}\\
\mathrm{B} + \mathrm{D} + \mathrm{E} & \stackrel{k_2}{\longleftrightarrow} & \mathrm{B}+\mathrm{G}+\mathrm{F}\\
\mathrm{F} + \mathrm{B} & \stackrel{k_3}{\longrightarrow} & \mathrm{H} + \mathrm{I}
\end{array}\right\}
\label{reactions}
\end{equation}
where $k_1$, $k_2$, and $k_3$ denote the chemical reaction rates. The second line denotes a reversible reaction and thus $k_2$ is split into $k_2^{-}$ and $k_2^{+}$, for the forward and backward reactions, respectively.

Let $[A](t) = A(t)/V$ be the concentration (in mol/litre) at time $t$ of A, let $\dot{A}(t) = \mathrm{d}[A](t)/\mathrm{d}t$ be the corresponding time derivative, and assume similar definitions for the remaining substances; B to I. Chemical reactions~(\ref{reactions}) lead to the following system of non-linear ODEs (suppressing the dependence on $t$):
\begin{equation}
\left.
\begin{array}{lll}
\dot{A} & = & - k_1 [A]\\
\dot{B} & = & 2k_1 [A] - k_3 [F][B]\\
\dot{C} & = & k_1 [A]\\
\dot{D} & = & - k_2^-[B][D][E] + k_2^+ [B][G][F]\\
\dot{E} & = & - k_2^-[B][D][E] + k_2^+ [B][G][F]\\
\dot{F} & = & k_2^- [B][D][E] - k_2^+ [B][G][F] - k_3 [F][B]\\
\dot{G} & = & k_2^-[B][D][E] - k_2^+ [B][G][F]\\
\dot{H} & = & k_3 [F][B]\\
\dot{I} & = & k_3 [F][B]
\end{array} \right\} \mbox{ for } t \in \mathcal{T} = [0,3000]\mbox{ seconds.}
\label{odes}
\end{equation}
The initial concentrations of all nine substances are zero, except for A, D and E which are $[A](0)=A_0/V$, $[D](0)=D_0/V$ and $[E](0)=E_0/V$, respectively. 

For given chemical reaction rates, the solution $\left([A](t),\dots,[I](t)\right)$ from~(\ref{odes}) provides the concentrations of the nine substances for given combinations of the controllable factors, $\mathbf{x}$, and reaction rates. The amounts of the nine substances are then given by $\left(A(t),\dots,I(t)\right) = V \times \left([A](t),\dots,[I](t)\right)$. However, the solution is analytically intractable and hence we employ a computationally expensive numerical method to find an approximate solution. In particular, we use the variable coefficient ordinary differential equation solver \citep{brown1989} as implemented in the \texttt{R} \citep{RRR} package \texttt{deSolve()} function \citep{sps2010}. However the methodology described in Section~\ref{METH} can be used in conjunction with any numerical method.

The first reaction is assumed to be instantaneous, which the chemical engineers model by assuming $k_1=10000$. The remaining reaction rates, $k_2^-$, $k_2^+$ and $k_3$, are unknown, and the temperature dependence is incorporated into the system of ODEs via the Arrehnius equation (see, for example, \citealt{L1984}). Hence, the general form for a reaction rate is
$$k_i = k_i^{(r)} \exp \left( \frac{E_i}{G} \left(\frac{1}{\lambda} - \frac{1}{\lambda^{(r)}}\right) \right)\,,\quad i=2,3\,, $$
where $\lambda^{(r)}$ is the reference temperature (here 298.15K), $k_i^{(r)}>0$ is the reaction rate at the reference temperature, $E_i>0$ is the activation energy and $G = 8.31445\mbox{ Jmol}^{-1}\mbox{K}^{-1}$ is the gas constant. By definition, the activation energies for the reversible reactions are equal, i.e. $E_2^+ = E_2^- = E_2$. This means there are $p=5$ unknown physical parameters, $\boldsymbol{\gamma} = (\gamma_1,\ldots,\gamma_5) = \left(k_2^{(r)-},k_2^{(r)+},k_3^{(r)},E_2,E_3\right)$. To increase computational efficiency of the methodology described in Section~\ref{METH}, we operate on the logarithm scale, i.e. consider $\theta_k = \log \gamma_k$, for $k=1,\dots,p$, and let $\boldsymbol{\theta} = \left(\theta_1,\dots,\theta_p\right) \in \Theta = \mathbb{R}^p$.

The quality control question of interest is to make choices of initial amounts of A, D and E, temperature, volume and time such that the amount of the three substances of interest satisfy the following specification limits (in mols):
\begin{equation}
\begin{array}{lll}
E(t) &<& 3\,,\\
F(t) &>& 20\,,\\
H(t) &<&3\,.
\end{array}
\label{constraints}
\end{equation}
Recall that F is the pharmaceutical product and its amount needs to be sufficiently large to make the production process economically viable. However E and H represent unwanted by-products with upper limits for safety reasons.

\subsection{Experiment and Physical Model}
We estimate the values of the unknown physical parameters $\boldsymbol{\theta}$ using observations from a small experiment. The chemical reactions~(\ref{reactions}) are observed for $I'=6$ combinations of the controllable factors $\mathbf{x}$, given in Table~\ref{Runs}. For run $i=1,\dots,I'$, we observe the amounts of E, F and H at a series of $n_i$ times $t_{i1},\dots,t_{in_i}$. The times range from 0.5 to 2902 seconds, with $n_i$ ranging from 17 to 20 time points per run. In total, there are $n=\sum_{i=1}^{I'} n_i =109$ observations of the amounts over all $I'$ runs. Note how run 6 is a repetition of run 5, thus there are $I=I'-1=5$ unique treatments. Runs 5 and 6 have observations taken at 18 different time points. However the last two time points for run 5 are at 1620 and 1825 seconds, whereas the last two time points for run 6 are taken four seconds later. Therefore there are $m=93$ unique combinations of controllable factors and time points. Let $\left(\mathbf{x}'_j,t_j\right)$ be the values of the controllable factors for each of these unique combinations ($j=1,\dots,m$). The experiment was not designed to be statistically optimal for either the estimation of parameters $\boldsymbol{\theta}$ or prediction for new treatments. 

\begin{table}[ht]
\caption{The six combinations of the controllable factors in the experiment.}
\label{Runs}
\begin{center}
\begin{tabular}{lllllcc} \hline
		& & \multicolumn{3}{c}{Initial amount (in mols)} & & \\
Run & Symbol & $A_0$ & $D_0$ & $E_0$ & Temperature ($\lambda$, Kelvin) & Volume ($V$, litres) \\ \hline
1 & $\mathbf{x}_1$ & 22.50 & 91.59 & 26.47 & 298.15 & 31.31 \\
2 & $\mathbf{x}_2$ & 45.00 & 91.59 & 26.47 &  298.15 & 32.56 \\
3 & $\mathbf{x}_3$ & 22.50 & 91.50 & 26.45 &  313.15 & 31.28 \\
4 & $\mathbf{x}_4$ & 45.00 & 91.50 & 26.45 & 313.15 & 32.53 \\
5 & $\mathbf{x}_5$ & 33.75 & 91.40 & 26.42 & 305.65 & 31.88 \\
6 & $\mathbf{x}_6$ & 33.75 & 91.40 & 26.42 & 305.65 & 31.88 \\ \hline
\end{tabular}
\end{center}
\end{table}

Let $K=3$ be the number of substances of interest. We denote by $\boldsymbol{\mu}\left(\boldsymbol{\theta};\mathbf{x},t\right) = \log\left(E(t),F(t),H(t)\right)$ the $K \times 1$ vector giving the log of the theoretical amount of the three substances of interest (E, F and H) obtained from the numerical solution to the system of ODEs at $\mathbf{x}$, time $t$ and parameters $\boldsymbol{\theta}$. 

Let $\mathbf{Y}_i$ be the $n_i \times K$ matrix containing the log observed amounts of E, F and H for the $i$th run, for $i=1,\dots,I'$. The $j$th row of $\mathbf{Y}_i$, denoted by $\mathbf{y}_{ij}$ is a $K \times 1$ containing the observed amounts for the $j$th time point of the $i$th run, for $i=1,\dots,I'$ and $j=1,\dots,n_i$. Let $\mathbf{Y}$ be the $n \times K$ matrix formed by stacking the matrices $\mathbf{Y}_1,\dots,\mathbf{Y}_{I'}$, i.e.
$$\mathbf{Y} = \left(\begin{array}{c}
\mathbf{Y}_1\\
\vdots\\
\mathbf{Y}_{I'} \end{array}\right).$$

The experimental apparatus cannot detect substance amounts lower than $\chi=0.01$ mols. Therefore, we have censored values when the observed amount is less than $\chi$. In total, there are zero censored observations in the first column of $\mathbf{Y}$, and 5 and 45 in the second and third columns, respectively. We take account of the censoring in our statistical modelling strategy described in Section~\ref{StatMod}.

\subsection{Statistical Model} \label{sec:model}
\label{StatMod}
We assume that
\begin{equation}
\mathbf{Y} = \mathbf{G} \left(\mathbf{M}(\boldsymbol{\theta}) + \mathbf{D}\right) + \mathbf{E}.
\label{EQN1}
\end{equation}
In (\ref{EQN1}), $\mathbf{M}(\boldsymbol{\theta})$ is the $m \times K$ matrix of unique model solutions, given by 
$$\mathbf{M}(\boldsymbol{\theta}) = \left(\begin{array}{c}
\mathbf{M}_1(\boldsymbol{\theta})\\
\vdots\\
\mathbf{M}_I(\boldsymbol{\theta}) \end{array}\right),$$
where $\mathbf{M}_i(\boldsymbol{\theta})$, for $i=1,\dots,I-1$, is the $n_i \times K$ matrix with $j$th row given by $\boldsymbol{\mu}(\boldsymbol{\theta};\mathbf{x}_i,t_{ij})$, and $\mathbf{M}_I(\boldsymbol{\theta})$ is the $(n_I+2) \times k$ matrix where the $j$th row is given by $\mu(\boldsymbol{\theta};\mathbf{x}_I,t_{Ij})$, for $j=1,\dots,n_I$, and the $(n_I+1)$th and $(n_I+2)$th rows are given by $\mu(\boldsymbol{\theta};\mathbf{x}_I,1624)$ and $\mu(\boldsymbol{\theta};\mathbf{x}_I,1829)$, respectively (corresponding to the two time points that differ between runs 5 and 6). 

Furthermore, $\mathbf{D}$ is the $m \times K$ matrix of model discrepancy errors given by
$$\mathbf{D} = \left(\begin{array}{c}
\mathbf{D}_1\\
\vdots\\
\mathbf{D}_I \end{array}\right),$$
which represents the discrepancy between the physical model and the true mean value of the process \citep{KOH2001}.  The $n \times m$ binary incidence matrix $\mathbf{G}$ identifies the rows of $\mathbf{M}+\mathbf{D}$ corresponding to the rows of $\mathbf{Y}$. Finally, $\mathbf{E}$ is the $n \times K$ matrix of observational errors.

The distribution of $\mathbf{E}$ is assumed to be
$$\mathbf{E} \sim \mathrm{MN}\left(\mathbf{0},\boldsymbol{\Omega},\mathbf{T}\right),$$
where $\mathrm{MN}$ denotes the matrix normal distribution \citep[e.g.][]{dawid1981}, $\boldsymbol{\Omega}$ is an unknown $K \times K$ column covariance matrix, and $\mathbf{T}$ is an $n \times n$ row correlation matrix. The matrix $\mathbf{T} = \mathrm{diag}\left\{\mathbf{T}_1,\dots,\mathbf{T}_{I'}\right\}$ is block-diagonal, with $jl$th element of $\mathbf{T}_i$ given by
$$T_{i,jl} = \exp \left( - \rho \left(t_{ij} - t_{il}\right)^2 \right)\qquad j,l=1,\ldots,n_i\,,$$
with time correlation parameter $\rho>0$ assumed unknown. This covariance structure for $\mathbf{E}$ assumes that the correlation between observations from the same run is dependent on the difference in time, but observations from different runs are independent.

We complete the model specification by specifying prior distributions for the discrepancy $\mathbf{D}$ and the model parameters. The discrepancy is assumed to follow a matrix normal distribution,
\begin{equation}
\mathbf{D} \sim \mathrm{MN}\left(\mathbf{0},\boldsymbol{\Sigma},\mathbf{S}\right)\,,
\label{EQN_D}
\end{equation}
where $\boldsymbol{\Sigma}$ is an unknown $K \times K$ covariance matrix and $\mathbf{S}$ is an $m \times m$ correlation matrix. The $jl$th element of $\mathbf{S}$ is given by
\begin{equation}
S_{jl} = \exp \left( - \psi_1 \sum_{f=1}^F \left(d_{jp} - d_{lp}\right)^2 - \psi_2 \left(t_j - t_l \right)^2\right),
\label{Aformula}
\end{equation}
for $j,l=1,\dots,m$ and where the model discrepancy correlation parameters $\psi_1>0$ and $\psi_2>0$ are unknown. In~(\ref{Aformula}), $d_{jf} = \left(x_{jf}' - x_{f,min}\right)/\left(x_{f,max} - x_{f,min}\right)$ is the scaled value of the $f$th controllable factor, for $f=1,\dots,F$ and $j=1,\dots,m$, where $x_{jf}'$ is the $f$th element of $\mathbf{x}'_j$. This correlation structure allows the differences in model discrepancy between different treatments to depend on the \textquotedblleft distance\textquotedblright \space between the controllable factors and time points for each treatment.

The chemical engineers have some limited prior knowledge on the location of $\boldsymbol{\theta}$. It is believed that the reaction rates are likely to lie between $10^{-8}$ and $10^{-4}$ and the activation energies between $10^2$ and $10^6$.  We encode this information by assuming that the reaction rates and activation energies have independent log-normal prior distributions with hyperparameters chosen so that there is probability 0.95 that the value of interest lies between the two limits above. This is achieved by setting the 0.025 and 0.975 quantiles to be the lower and upper limits respectively. In summary, we assume 
$$\log \boldsymbol{\theta} \sim \mathrm{N}\left(\boldsymbol{\mu},\boldsymbol{\Delta}\right)\,,$$
where $\boldsymbol{\mu}=\left(-13.8,-13.8,-13.8,9.21,9.21\right)^{\mathrm{T}}$ and $\boldsymbol{\Delta}= 5.52 \mathbf{I}_p$. There is no prior information available for the remaining parameters, so we specify vague prior distributions that contribute negligible information. The variance matrices $\boldsymbol{\Omega}$ and $\boldsymbol{\Sigma}$ are both assumed to have inverse-Wishart prior distributions with $\nu = 4$ degrees of freedom and an identity scale matrix \citep[][Chapter 3]{G2014}. The correlation parameters $\rho$, $\psi_1$ and $\psi_2$ are given independent exponential prior distributions with mean equal to one following the arguments of, for example, \cite{OW2016}. In Section~\ref{RES} we discuss the results of a sensitivity analysis to assess the robustness to the choice of these vague prior distributions.

Let $\mathbf{y}=\mathrm{vec}(\mathbf{Y})$, $\mathbf{m}(\boldsymbol{\theta})=\mathrm{vec}(\mathbf{M}(\boldsymbol{\theta}))$, $\mathbf{d}=\mathrm{vec}(\mathbf{D})$ and $\mathbf{e}=\mathrm{vec}(\mathbf{E})$, with the $\mathrm{vec}$ operator stacking columns of a matrix. Then a model specification equivalent to~(\ref{EQN1}) is given by
\begin{equation}
\mathbf{y} = \mathbf{H}\left(\mathbf{m}(\boldsymbol{\theta}) + \mathbf{d}\right) + \mathbf{e}\,,
\label{EQN1_a}
\end{equation}
where $\mathbf{H}= \mathbf{I}_k \otimes \mathbf{G}$, 
\begin{eqnarray}
\mathbf{e} & \sim & \mathrm{N}\left(\mathbf{0},\boldsymbol{\Omega} \otimes \mathbf{T}\right)\,,\label{EQN1_x}\\
\mathbf{d} & \sim & \mathrm{N}\left(\mathbf{0},\boldsymbol{\Sigma} \otimes \mathbf{S}\right)\,,\label{EQN1_y}
\end{eqnarray}
and $\otimes$ denotes the Kronecker product.

Let $\mathbf{y}_S$ and $\mathbf{y}_C$ denote the elements of $\mathbf{y}$ which are fully observed (i.e. greater than $\log \chi$) and censored (i.e. less than $\log \chi$), respectively. Therefore, we need to evaluate the posterior distribution (conditional on $\mathbf{y}_S$) of the model parameters and censored observations. This distribution is given by
\begin{equation}
\pi(\boldsymbol{\theta},\boldsymbol{\Omega},\rho,\mathbf{d},\boldsymbol{\Sigma},\boldsymbol{\psi},\mathbf{y}_C|\mathbf{y}_S) \propto \pi(\mathbf{y}|\boldsymbol{\theta},\mathbf{d},\boldsymbol{\Omega},\rho)\pi(\mathbf{d}|\boldsymbol{\Sigma},\boldsymbol{\psi}) \pi(\boldsymbol{\theta})\pi(\boldsymbol{\Omega})\pi(\rho)\pi(\boldsymbol{\Sigma})\pi(\boldsymbol{\psi})\,,
\label{finalpost}
\end{equation}
where $\pi(\mathbf{y}|\boldsymbol{\theta},\mathbf{d},\boldsymbol{\Omega},\rho)$ is the complete data likelihood given by~(\ref{EQN1_a}) and (\ref{EQN1_x}), $\pi(\mathbf{d}|\boldsymbol{\Sigma},\boldsymbol{\psi})$ is the density of the vectorised model discrepancy~(\ref{EQN1_y}), and $\pi(\boldsymbol{\theta})$, $\pi(\boldsymbol{\Omega})$, $\pi(\rho)$, $\pi(\boldsymbol{\Sigma})$ and $\pi(\boldsymbol{\psi})$ are prior densities for the model parameters.

\subsection{Prediction}
\label{Pred}

Let $\mathbf{y}_0$ be a $K \times 1$ vector denoting the predicted log amounts of E, F and H for arbitary controllable factors $\mathbf{x}_0 \in \mathcal{X}$ and $t_0 \in \mathcal{T}$. The posterior predictive distribution of $\mathbf{y}_0$ is given by
\begin{equation}
\pi(\mathbf{y}_0|\mathbf{y}_S) = \int \pi(\mathbf{y}_0|\boldsymbol{\theta},\mathbf{D},\boldsymbol{\psi},\boldsymbol{\Omega},\boldsymbol{\Sigma}) \pi(\boldsymbol{\theta},\mathbf{D},\boldsymbol{\psi},\boldsymbol{\Omega},\boldsymbol{\Sigma}|\mathbf{y}_S) \mathrm{d}\boldsymbol{\theta}\mathrm{d}\mathbf{D}\mathrm{d}\boldsymbol{\psi}\mathrm{d}\boldsymbol{\Omega}\mathrm{d}\boldsymbol{\Sigma}\,,
\label{postpred}
\end{equation}
where
$$\mathbf{y}_0|\boldsymbol{\theta},\mathbf{D},\boldsymbol{\psi},\boldsymbol{\Omega},\boldsymbol{\Sigma} \sim \mathrm{N}\left(\boldsymbol{\mu}(\boldsymbol{\theta};\mathbf{x},t) + \mathbf{D}^{\rm T}\mathbf{S}^{-1}\mathbf{s}, \left(1-\mathbf{s}^{\rm T}\mathbf{S}^{-1}\mathbf{s}\right)\boldsymbol{\Sigma} + \boldsymbol{\Omega}\right)\,,$$
and $\mathbf{s}$ is an $m \times 1$ vector with $j$th element given by
$$s_j = \exp \left( - \psi_1 \sum_{f=1}^F \left(d_{jf} - d_{0f}\right)^2 - \psi_2\left(t_j - t_0\right)^2 \right)\,,$$
with $d_{0f} = \left(x_{0f} - x_{f,min}\right)/\left(x_{f,max} - x_{f,min}\right)$ being the scaled value of each element of $\mathbf{x}_0$, for $f=1,\dots,F$. 

The probability of E, F and H satisfying the specification limits (\ref{constraints}) at point $(\mathbf{x}_0, t_0)$ is given by 
$$\mathrm{P}(\mathbf{y}_0 \in \mathcal{Y}|\mathbf{y}_S) = \int_{\mathcal{Y}} \pi(\mathbf{y}_0|\mathbf{y}_S) \mathrm{d}\mathbf{y}_0\,,,$$
where $\mathcal{Y} = \left\{ \boldsymbol{\eta} \in \mathbb{R}^K|\eta_1 < \log 3, \eta_2 > \log 20, \eta_3 < \log 3 \right\}$.



\section{Methodology}
\label{METH}

The posterior and predictive distributions given by~(\ref{finalpost}) and~(\ref{postpred}), respectively, will be analytically intractable. Our general approach to numerically approximating these distributions uses two phases. The \textbf{Sampling Phase} involves generating a sample from the posterior distribution using MCMC methods. The \textbf{Prediction Phase} uses this MCMC sample to generate a sample from the posterior predictive distribution and thus estimate the probability $\mathrm{P}(\mathbf{y}_0 \in \mathcal{Y}|\mathbf{y}_S)$ by the proportion of sampled values satisfying the specification limits. These phases make use of Gibbs sampling and parallel tempering algorithms, and multivariate Gaussian process emulators for the numerical solution to the ODEs.

\subsection{Gibbs sampling and parallel tempering}
\label{GS}
We use Gibbs sampling in conjunction with parallel tempering to generate a sample from the joint posterior distribution~\eqref{finalpost} of model parameters and censored observations. To improve the efficiency of the Gibbs sampler, we employ hierarchical centring \citep{PRS2003} and reparameterise model~(\ref{EQN1_a}) as
\begin{equation}
\mathbf{y} = \mathbf{H}\mathbf{d}^* +  \mathbf{e}\,,
\label{EQN1_b}
\end{equation}
where 
\begin{equation}
\mathbf{d}^* \sim \mathrm{N}\left(\mathbf{m}(\boldsymbol{\theta}),\boldsymbol{\Sigma} \otimes \mathbf{S} \right)\,.
\label{EQN2_b}
\end{equation}
Sampled values of $\mathbf{d}^*$ can easily be transformed to values of $\mathbf{d}$ using $\mathbf{d} = \mathbf{d}^* - \mathbf{m}(\boldsymbol{\theta})$. The posterior distribution of model parameters and censored observations is now given by
$$\pi(\boldsymbol{\theta},\boldsymbol{\Omega},\rho,\mathbf{d}^*,\boldsymbol{\Sigma},\boldsymbol{\psi},\mathbf{y}_C|\mathbf{y}_S) \propto \pi(\mathbf{y}|\mathbf{d}^*,\boldsymbol{\Omega},\rho)\pi(\mathbf{d}^*|\boldsymbol{\theta},\boldsymbol{\Sigma},\boldsymbol{\psi})\pi(\boldsymbol{\theta})\pi(\boldsymbol{\Omega})\pi(\rho)\pi(\boldsymbol{\Sigma})\pi(\boldsymbol{\psi})\,,$$
where $\pi(\mathbf{y}|\mathbf{d}^*,\boldsymbol{\Omega},\rho)$ is the complete-data likelihood defined by~(\ref{EQN1_b}) and $\pi(\mathbf{d}^*|\boldsymbol{\theta},\boldsymbol{\Sigma},\boldsymbol{\psi})$ is defined by~(\ref{EQN2_b}). The full conditional posterior distributions of $\mathbf{d}^*$, $\boldsymbol{\Omega}$ and $\boldsymbol{\Sigma}$ are of known form and are given in Appendix~\ref{APP_a}. The full conditional posterior distribution of $\mathbf{y}_C$ is a truncated multivariate normal distribution (see Appendix~\ref{APP_a}) and there exist efficient methods for generating values from such distributions \citep{G1991}. 

For the remaining parameters $\boldsymbol{\theta}$, $\rho$ and $\boldsymbol{\psi}$, the full conditional distributions are not of known form, so a Metropolis-within-Gibbs step is employed. Specifically, we exploit the Riemann manifold Langevin Metropolis-Hastings (RMLMH) algorithm \citep{GC2011} . This method uses derivative information on the posterior surface to construct an efficient proposal distribution. The algorithmic details are briefly described in Appendix~\ref{APP_b} and the components necessary for its application to the full conditional posterior distributions of $\boldsymbol{\theta}$, $\rho$ and $\boldsymbol{\psi}$ are given in Appendix~\ref{APP_c}. 

Parallel tempering \citep[see, for example,][pp. 299-300]{G2014} is used to improve the efficiency of sampling from a potentially multi-modal posterior distribution. It involves defining $R$ distributions by a series of increasing ``temperatures'', where the lowest temperature distribution corresponds to the posterior distribution. An MCMC chain is initialized under each distribution, and the parallel tempering scheme allows for swaps between the states of the chains for distributions at different temperatures. The lowest temperature MCMC chain is a sample from the posterior distribution. Parallel tempering allows larger moves (e.g. between different areas of high posterior density) made at the higher temperature chains to be passed down to the lower temperature chains, hence improving mixing. 

Let $\boldsymbol{\delta} = \left(\boldsymbol{\theta},\boldsymbol{\Omega},\rho,\mathbf{d}^*,\boldsymbol{\Sigma},\boldsymbol{\psi},\mathbf{y}_C\right)$ be the collection of unknown parameters and censored observations. For $r = 1, \dots, R$, define the distribution for the $r$th chain to have density proportional to
$$U_r(\boldsymbol{\delta}) = \exp \left( \frac{ \log \pi(\mathbf{y}_S|\boldsymbol{\delta})+\log\pi(\boldsymbol{\delta})}{\tau_r}\right),$$
where $1 = \tau_1 < \dots < \tau_R$ are a series of temperatures. A parallel tempering iteration involves applying either i) a sampling step (with probability $\omega$); or ii) a swap step (with probability $1-\omega$). In the sampling step, the current values $\boldsymbol{\delta}_{(r)}^c$ in the $r$th chain are updated to $\boldsymbol{\delta}_{(r)}^{c+1}$, for $r=1,\dots,R$, using the Gibbs sampling algorithm. In the swap step, two neighbouring chains ($r$ and $r+1$) are chosen at random. With probability
$$\min \left[ 1, \frac{U_{r+1}(\boldsymbol{\delta}_{(r)}^c) U_{r}(\boldsymbol{\delta}^c_{(r+1)})}{U_{r+1}(\boldsymbol{\delta}^c_{(r+1)}) U_{r}(\boldsymbol{\delta}^c_{(r)})}\right]\,,$$
set $\boldsymbol{\delta}_{(r)}^{c+1} = \boldsymbol{\delta}_{(r+1)}^c$ and $\boldsymbol{\delta}_{(r+1)}^{c+1} = \boldsymbol{\delta}_{(r)}^c$. 
To complete sampling step (i) for each distribution at temperature $t_r$, the full conditional and RMLMH components need to be adjusted (see Appendices~\ref{APP_a} and~\ref{APP_b} for details).

\subsection{Amended Gibbs sampling and parallel tempering} \label{compsav}

One iteration of the RMLMH step in the Gibbs sampling algorithm outlined in Section~\ref{GS} requires one evaluation of $\mathbf{m}(\boldsymbol{\theta})$ to calculate the acceptance probability and one evaluation each of sensitivities $\partial \mathbf{m}(\boldsymbol{\theta})/ \partial \boldsymbol{\theta}$ and $\partial^2 \mathbf{m}(\boldsymbol{\theta})/ \partial \boldsymbol{\theta} \partial \theta_k$ to form the proposal distribution. To evaluate $\mathbf{m}(\boldsymbol{\theta})$ we are required to evaluate $\boldsymbol{\mu}(\boldsymbol{\theta};\mathbf{x}_i,t_{ij})$, for $i=1,\dots,I$ and $j=1,\dots,n_i$. The derivative terms require evaluation of $\partial \boldsymbol{\mu}(\boldsymbol{\theta};\mathbf{x}_i,t_{ij})/\partial \boldsymbol{\theta}$ and $\partial^2 \boldsymbol{\mu}(\boldsymbol{\theta};\mathbf{x}_i,t_{ij})/\partial \boldsymbol{\theta} \partial \theta_k$, for $i=1,\dots,I$, $j=1,\dots,n_i$ and $k=1,\dots,p$. 

The system of ODEs can be augmented with additional equations to numerically solve for the first- and second-order sensitivities \citep{VV1984,GC2011}. However this will substantially increase the computational burden of the numerical solution. Note that to generate a sample of size $B$ using parallel tempering with $R$ chains will require $BR$ evaluations of the three functions for each $\mathbf{x}_i$ and $t_{ij}$, for $i=1,\dots,I$, $j=1,\dots,n_i$ and $k=1,\dots,p$. Once we have generated a sample of size $B$ from the posterior distribution, to estimate the probability $\mathrm{P}(\mathbf{y}_0 \in \mathcal{Y}|\mathbf{y}_S)$, we require $B$ further evaluations of $\boldsymbol{\mu}(\boldsymbol{\theta};\mathbf{x},t)$, for each $\mathbf{x} \in \mathcal{X}$ and $t \in \mathcal{T}$ at which we wish to evaluate this probability.

Making such a large number of evaluations of the numerical solution is clearly computationally infeasible. Instead we form an approximation to the numerical solution by the construction of a statistical emulator through a computer experiment \citep{sacks1989}. The numerical solution to the ODEs is evaluated at a selection of $N$ combinations of parameters $\boldsymbol{\theta}$; controllable variables $\mathbf{x}$; and times $t$. These evaluations are treated as data to which a statistical model, termed an emulator, is fitted. The emulator is essentially a predictive distribution (denoted by $\mathrm{Q}(\boldsymbol{\theta},\mathbf{x},t)$) from which fast predictions of the numerical solution can be obtained. For example, the predictive mean, $\hat{\boldsymbol{\mu}}(\boldsymbol{\theta};\mathbf{x},t)$, will be used as a point prediction. For more details on computer experiments and statistical emulators, see, for example, \cite{SWN2003}, \citet{fang2006} and \citep[][Section V]{dean2015}.

In the Sampling Phase, we apply the Gibbs sampling and parallel tempering algorithm with the following amendments:
\begin{enumerate}
\item
All evaluations of sensitivities $\partial \boldsymbol{\mu}(\boldsymbol{\theta};\mathbf{x},t)/\partial \boldsymbol{\theta}$ and $\partial^2 \boldsymbol{\mu}(\boldsymbol{\theta};\mathbf{x},t)/\partial \boldsymbol{\theta} \partial \theta_k$ are replaced by $\partial \hat{\boldsymbol{\mu}}(\boldsymbol{\theta};\mathbf{x},t)/\partial \boldsymbol{\theta}$ and $\partial^2 \hat{\boldsymbol{\mu}}(\boldsymbol{\theta};\mathbf{x},t)/\partial \boldsymbol{\theta} \partial \theta_k$, respectively, for $k=1,\dots,p$.
\item
$R+1$ parallel chains are constructed for temperatures $1=\tau_0=\tau_1 < \tau_2 < \dots < \tau_R$. In chains $r \ge 1$, all evaluations of $\boldsymbol{\mu}(\boldsymbol{\theta};\mathbf{x},t)$ in the evaluation of $U_r(\boldsymbol{\delta})$ in the acceptance probability are replaced by evaluation of $\hat{\boldsymbol{\mu}}(\boldsymbol{\theta};\mathbf{x},t)$. For evaluation of $U_0(\boldsymbol{\delta})$ in chain $r=0$ with temperature $\tau_0=1$, evaluation of the actual $\boldsymbol{\mu}(\boldsymbol{\theta};\mathbf{x},t)$ is retained.
\end{enumerate}
This modifications results in the sample generated in chain $r=0$ being from the true posterior distribution of $\boldsymbol{\delta}$ but the overall sampling scheme requiring only $B$ evaluations of $\boldsymbol{\mu}(\boldsymbol{\theta};\mathbf{x}_i,t_{ij})$ for $i=1,\dots,I$ and $j=1,\dots,n_i$.

In the Prediction Phase, to estimate the probability of satisfying the specification limits, all evaluations of $\boldsymbol{\mu}(\boldsymbol{\theta};\mathbf{x},t)$ are replaced by a value generated from $\mathrm{Q}(\boldsymbol{\theta},\mathbf{x},t)$. Sampling from this predictive distribution means the estimate of the probability takes account of additional uncertainty introduced by using an emulator approximation.


\subsection{Multivariate Gaussian process emulator} \label{MGP}
In this paper, we employ the multivariate Gaussian process (MGP; \citealt{COH2010}) emulator. The numerical solution to the ODEs, $\boldsymbol{\mu}(\boldsymbol{\theta};\mathbf{x},t)$, is evaluated at each of the elements of the set (or meta-design) $\zeta = \left\{ \left(\boldsymbol{\theta}^*_1,\mathbf{x}^*_1,t_1^*\right),\dots,\left(\boldsymbol{\theta}^*_N,\mathbf{x}_N^*,t_N^*\right)\right\}$. The superscript $^*$ notation has been introduced to distinguish the meta-design from the design of the physical experiment. Let $\mathbf{z}_i = \boldsymbol{\mu}(\boldsymbol{\theta}^*_i;\mathbf{x}^*_i,t^*_i)$, $\mathbf{Z}$ be the $N \times K$ matrix with $i$th row given by $\mathbf{z}_i^{\mathrm{T}}$, and $\mathbf{w}_i = \left(\boldsymbol{\theta}^*_i;\mathbf{x}^*_i,t^*_i\right)$ be the $U \times 1$ vector of all inputs, where $U= p + F + 1 = 11$. The central assumption of the MGP is that $\mathbf{Z}$ follows a matrix normal distribution,
$$\mathbf{Z} \sim \mathrm{MN}\left(\mathbf{1}_N \boldsymbol{\beta}^{\rm T},\boldsymbol{\Phi},\mathbf{P}\right)\,,$$
where $\boldsymbol{\beta}$ is a $K \times 1$ vector of common column means of $\mathbf{Z}$, $\boldsymbol{\Phi}$ is a $K \times K$ unstructured column covariance matrix, $\mathbf{P}$ is an $N \times N$ row correlation matrix and $\mathbf{1}_N$ is the $N \times 1$ vector of ones. We model the $ij$th element of $\mathbf{P}$ as
$$P_{ij} = \exp \left( - \sum_{u=1}^{U} \alpha_u \left(w_{iu} - w_{ju}\right)^2 \right)\,,$$
where $w_{iu}$ is the $u$th element of $\mathbf{w}_i$ and $\alpha_u>0$ are correlation parameters, for $u=1,\dots,U$. 

Suppose we wish to predict $\mathbf{z}_0 = \boldsymbol{\mu}(\boldsymbol{\theta}_0;\mathbf{x}_0,t_0)$. Let 
\begin{eqnarray}
\hat{\boldsymbol{\beta}} & = & \frac{\mathbf{1}_N^{\rm T} \mathbf{P}^{-1} \mathbf{Z}}{\mathbf{1}_N^{\rm T} \mathbf{P}^{-1} \mathbf{1}_N}\,,\label{eqn11}\\
\hat{\boldsymbol{\Psi}} & = & \mathbf{Z}^{\rm T} \mathbf{P}^{-1}\left(\mathbf{I}_N - \frac{\mathbf{1}_N\mathbf{1}_N^{\rm T} \mathbf{P}^{-1}}{\mathbf{1}_N^{\rm T} \mathbf{P}^{-1}\mathbf{1}_N} \right)\mathbf{Z}\,, \label{eqn12}
\end{eqnarray}
and $\mathbf{w}_0 = \left(\boldsymbol{\theta}_0,\mathbf{x}_0,t_0\right)$, with $\mathbf{I}_N$ the $N \times N$ identity matrix.

It can be shown that the predictive distribution of $\mathbf{z}_0$, i.e. $\mathrm{Q}(\boldsymbol{\theta},\mathbf{x},t)$, conditional on outputs $\mathbf{z}_1,\ldots,\mathbf{z}_N$ is given by
\begin{equation}
\mathrm{N}\left(\hat{\boldsymbol{\mu}}(\boldsymbol{\theta};\mathbf{x},t),\hat{\boldsymbol{\Pi}}(\boldsymbol{\theta};\mathbf{x},t)\right)\,,
\label{emul}
\end{equation}
where
\begin{eqnarray}
\hat{\boldsymbol{\mu}}(\boldsymbol{\theta};\mathbf{x},t) & = & \hat{\boldsymbol{\beta}} + \mathbf{p}_0^{\mathrm{T}} \mathbf{P}^{-1}  \left(\mathbf{Z} - \mathbf{1}_N \hat{\boldsymbol{\beta}}^{\rm T} \right)\,,\label{eqn13}\\
\hat{\boldsymbol{\Pi}}(\boldsymbol{\theta};\mathbf{x},t) & = & \left( 1 - \mathbf{p}_0^{\mathrm{T}} \mathbf{P}^{-1}\mathbf{p}_0\right) \hat{\boldsymbol{\Psi}}\,, \label{eqn14}
\end{eqnarray}
with $N \times 1$ vector $\mathbf{p}_0$ having $i$th element
$$p_{0i} = \exp \left( - \sum_{u=1}^{U} \alpha_u \left(w_{iu} - w_{0u}\right)^2 \right),$$
and $w_{0u}$ being the $u$th element of $\mathbf{w}_0$. The predictive distribution given by (\ref{emul}) is conditional on the correlation parameters, $\boldsymbol{\alpha} = \left(\alpha_1,\dots,\alpha_{U}\right)$. We replace these values by their marginal posterior mode \citep[e.g.][]{OW2016} although a fully Bayesian approach could be taken with these parameters integrated out with respect to their posterior distribution, conditional on $\mathbf{Z}$.

\subsubsection{Meta-design for the Sampling and Prediction Phases} \label{expalg}

The two emulators formed prior to the Sampling and Prediction Phases depend on the choice of the meta-design $\zeta$. The numerical solution, $\boldsymbol{\mu}(\boldsymbol{\theta};\mathbf{x},t)$, has three different inputs: $\boldsymbol{\theta}$, $\mathbf{x}$ and $t$. We construct the meta-design as a cartesian product of designs for each of these three input types, 
$$\zeta = \zeta_1 \times \zeta_2 \times \zeta_3,$$
where $\zeta_1 = \left\{\boldsymbol{\theta}^*_1,\dots,\boldsymbol{\theta}^*_{N_1}\right\}$, $\zeta_2 = \left\{\mathbf{x}^*_1,\dots,\mathbf{x}^*_{N_2}\right\}$, $\zeta_3 = \left\{t^*_1,\dots,t^*_{N_3}\right\}$, and $N = \prod_{i=1}^M N_i$ where $M=3$ for our experiment. Separate designs in $t$ and $\boldsymbol{\theta}$ and $\mathbf{x}$ allow us to exploit computational efficiencies in the computation of the numerical solution to the ODEs; computing $\boldsymbol{\mu}(\boldsymbol{\theta};\mathbf{x},t_1)$ and $\boldsymbol{\mu}(\boldsymbol{\theta};\mathbf{x},t_2)$, for $t_1 \ne t_2$, requires a negligible increase in computational effort over just computing $\boldsymbol{\mu}(\boldsymbol{\theta};\mathbf{x},t_1)$.

This structure of meta-design also allows the emulator row correlation matrix, $\mathbf{P}$, to be decomposed as
$$\mathbf{P} = \bigotimes_{i=1}^M \mathbf{P}_i = \mathbf{P}_1 \otimes \mathbf{P}_2 \otimes \mathbf{P}_3,$$
where the $ij$th elements of $\mathbf{P}_1$, $\mathbf{P}_2$ and $\mathbf{P}_3$ are
$$
\begin{array}{lllr}
P_{1,ij} & = & \exp \left( - \sum_{u=1}^p \alpha_u \left(\theta^*_{iu} - \theta^*_{ju}\right)^2 \right) & \mbox{for $i,j = 1,\dots, N_1$}\,,\\
P_{2,ij} & = & \exp \left( - \sum_{u=1}^F \alpha_{p+u} \left(x^*_{iu} - x^*_{ju}\right)^2 \right) & \mbox{for $i,j = 1,\dots, N_2$}\,,\\
P_{3,ij} & = & \exp \left( - \alpha_U \left(t^*_i - t^*_j\right)^2 \right) & \mbox{for $i,j = 1,\dots, N_3$}\,,
\end{array}$$
respectively. Hence,
$$\mathbf{P}^{-1} = \bigotimes_{i=1}^M \mathbf{P}_i^{-1},$$
significantly reducing the computational burden of constructing the emulator and improving numerical stability. Similarly, for a meta-design with this structure, ~(\ref{eqn11}),~(\ref{eqn12}),~(\ref{eqn13}) and~(\ref{eqn14}) also simplify: 
\begin{eqnarray*}
\hat{\boldsymbol{\beta}} & = & \frac{\left( \bigotimes_{i=1}^M \mathbf{1}_{N_i}^{\rm T} \mathbf{P}_i^{-1}\right)\mathbf{Z}}{\prod_{i=1}^M \mathbf{1}_{N_i}^{\rm T} \mathbf{P}_i^{-1} \mathbf{1}_{N_i}},\\
\hat{\boldsymbol{\Psi}} & = & \mathbf{Z}^{\rm T} \left[ \bigotimes_{i=1}^M \mathbf{P}_i^{-1} -  \bigotimes_{i=1}^M \mathbf{P}_i^{-1}\mathbf{1}_{N_i}\mathbf{1}_{N_i}^{\rm T} \mathbf{P}_i^{-1} \right]\mathbf{Z},\\
\hat{\boldsymbol{\mu}}(\boldsymbol{\theta};\mathbf{x},t) & = & \hat{\boldsymbol{\beta}} + \left( \bigotimes_{i=1}^M \mathbf{p}_i^{\rm T} \mathbf{P}_i^{-1}\right) \left(\mathbf{Z} - \mathbf{1}_N \hat{\boldsymbol{\beta}}^{\rm T}\right),\\
\hat{\boldsymbol{\Pi}}(\boldsymbol{\theta};\mathbf{x},t) & = & \left( 1 - \prod_{i=1}^M \mathbf{p}_i^{\rm T} \mathbf{P}_i^{-1} \mathbf{p}_i\right) \hat{\boldsymbol{\Psi}},
\end{eqnarray*}
where vectors $\mathbf{p}_1$, $\mathbf{p}_2$ and $\mathbf{p}_3$ have $j$th elements given by
$$\begin{array}{lllr}
p_{1j} & = & \exp \left( - \sum_{u=1}^p \alpha_u \left(\theta^*_{ju} - \theta_u\right)^2\right),& \mbox{for $j=1,\dots,N_1$\,,}\\
p_{2j} & = & \exp \left( - \sum_{u=1}^p \alpha_{p+u} \left(x^*_{ju} - x_u\right)^2\right),& \mbox{for $j=1,\dots,N_2$\,,}\\
p_{3j} & = & \exp \left( - \alpha_U \left(t^*_j - t\right)^2 \right),& \mbox{for $j=1,\dots,N_3$\,,}
\end{array}$$  
respectively.


When we choose $\zeta$ to construct the emulator for the \textbf{Sampling Phase}, we choose the values of $\mathbf{x}$ and $t$ to coincide with the design of the physical experiment, i.e. to be equal to $\mathbf{x}_1,\dots,\mathbf{x}_I$ and $t_1,\dots,t_m$. Hence uncertainty in the prediction will only result from the different choice of parameter $\boldsymbol{\theta}$. For our experiment, this means fixing $\zeta_2 = \left\{\mathbf{x}_1,\dots,\mathbf{x}_5 \right\}$ and $\zeta_3 = \left\{t_1,\dots,t_m\right\}$, i.e. $N_2 = I = 5$ and $N_3 = m$. 

We use the following exploratory algorithm to adaptively choose $\zeta_1$. This uses a hybrid of the methodologies proposed by \citet{R2003}, \citet{FNL2011} and \citet{OW2013}. \citet{R2003} first proposed using MCMC methods to adaptively improve an emulator to an unnormalised posterior density based on a computationally expensive function in light of observed data. \citet{FNL2011} extended this methodology by using parallel tempering, and \citet{OW2013} proposed to emulate the computationally expensive function, as opposed to the posterior density, to make model criticism more efficient.

The following \textit{exploratory algorithm} iteratively improves the accuracy of the emulator in the region of the parameter space $\Theta$ corresponding to high posterior density.
\begin{enumerate}
\item
Generate the set $\zeta^0_1 = \left\{\boldsymbol{\theta}^*_1,\dots,\boldsymbol{\theta}^*_{N^0_1}\right\}$ as a sample of size $N^0_1$ from the prior distribution of $\boldsymbol{\theta}$.
\item
Let $\zeta^0 = \zeta^0_1 \times \zeta_2 \times \zeta_3$ and fit the MGP to the resulting evaluations from $\boldsymbol{\mu}(\boldsymbol{\theta};\mathbf{x},t)$.
\item
Let $\zeta = \zeta^0$ and repeat the following steps until $\boldsymbol{\mu}(\boldsymbol{\theta};\mathbf{x},t)$ as been evaluated a total of $N$ times.
		\begin{enumerate}
		\item
		Perform $L$ iterations of the Gibbs sampling and parallel tempering scheme (for chains $r=1,\dots,R$) where evaluation of $\boldsymbol{\mu}(\boldsymbol{\theta};\mathbf{x},t)$ is replaced by evaluation of $\hat{\boldsymbol{\mu}}(\boldsymbol{\theta};\mathbf{x},t)$ in all instances.
		\item
		Evaluate $\tilde{\mathbf{z}}_{rij} = \boldsymbol{\mu}(\boldsymbol{\theta}_{(r)};\mathbf{x}_i,t_j)$, for $r=1,\dots,R$, $i=1,\dots,I$ and $j=1,\dots,n_i$, where $\boldsymbol{\theta}_{(r)}$ is the current value of $\boldsymbol{\theta}$ in the $r$th chain. Augment the matrix $\mathbf{Z}$ by $\tilde{\mathbf{z}}_{rij}$ for $r=1,\dots,R$, $i=1,\dots,I$ and $j=1,\dots,n_i$. Refit the MGP.
		\end{enumerate}
\end{enumerate}

The meta-design for the \textbf{Prediction Phase} is constructed as the cartesian product of three space-filling designs. Firstly, $\zeta_1$ is constructed as a space-filling  \citep[e.g.][]{JMY1990} sub-sample of size $N_1$, selected using the \texttt{cover.design} function in the \texttt{R} package \texttt{Fields} \citep{fields} using the sample generated from the posterior distribution of $\boldsymbol{\theta}$ in the Sampling Phase as a candidate list. Secondly, $\zeta_2$ and $\zeta_3$ are chosen as Latin hypercube designs in $\mathcal{X}$ and $\mathcal{T}$ of sizes $N_2$ and $N_3$, respectively (\citealp{SWN2003}, Ch.~5). 

\section{Results}
\label{RES}

\subsection{Sampling Phase}

To create the meta-design for the Sampling Phase, we set $N^0_1 = 50$, $N_1=100$ and set up $R=5$ parallel chains. This results in ten iterations of the exploratory algorithm in Section~\ref{expalg} with $L=50$. We adapt the temperatures of the parallel chains using the methodology of \cite{MBV2013}. Figure~\ref{fig:PLOT1} shows the resulting values of $\boldsymbol{\theta}^*$ in meta-design $\zeta_1$.  It is clear that the exploratory algorithm selects points from a concentrated region of $\Theta$.

We then obtain a sample of size $B = 50000$ from the posterior distribution of $\boldsymbol{\delta}$ using the amended Gibbs sampling and parallel tempering algorithm presented in Section~\ref{compsav}. Convergence was assessed informally via trace plots (not shown) which showed that the chains had mixed adequately and each posterior was unimodal. Figure~\ref{fig:PLOT2} shows plots of the prior and estimated posterior densities for each element of $\boldsymbol{\theta}$ as well as the elements of $\boldsymbol{\psi}$ and $\rho$. Clearly the data has led to an increase in information from the prior to posterior distributions.


\begin{knitrout}
\definecolor{shadecolor}{rgb}{0.969, 0.969, 0.969}\color{fgcolor}\begin{figure}

{\centering \includegraphics[width=6in]{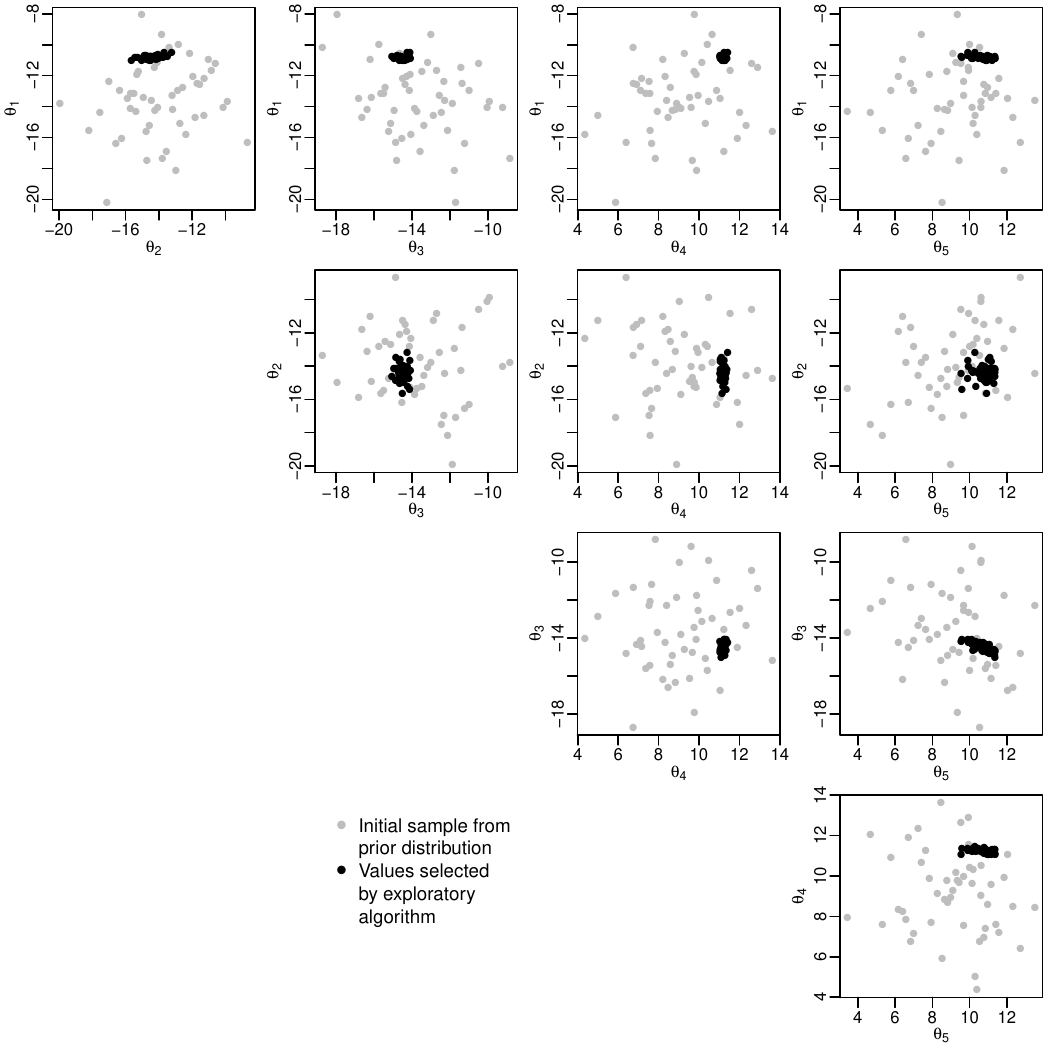} 

}

\caption[Values of $\boldsymbol{\theta}$ in the meta-design $\zeta_1$ for the Sampling Phase as found by the exploratory algorithm in Section~\ref{expalg}]{Values of $\boldsymbol{\theta}$ in the meta-design $\zeta_1$ for the Sampling Phase as found by the exploratory algorithm in Section~\ref{expalg}: grey points are the $N_1^0$ initial values generated from the prior distribution and black points are the values selected in step~3 of the exploratory algorithm.}\label{fig:PLOT1}
\end{figure}

\end{knitrout}

\begin{knitrout}
\definecolor{shadecolor}{rgb}{0.969, 0.969, 0.969}\color{fgcolor}\begin{figure}

{\centering \includegraphics[width=6in]{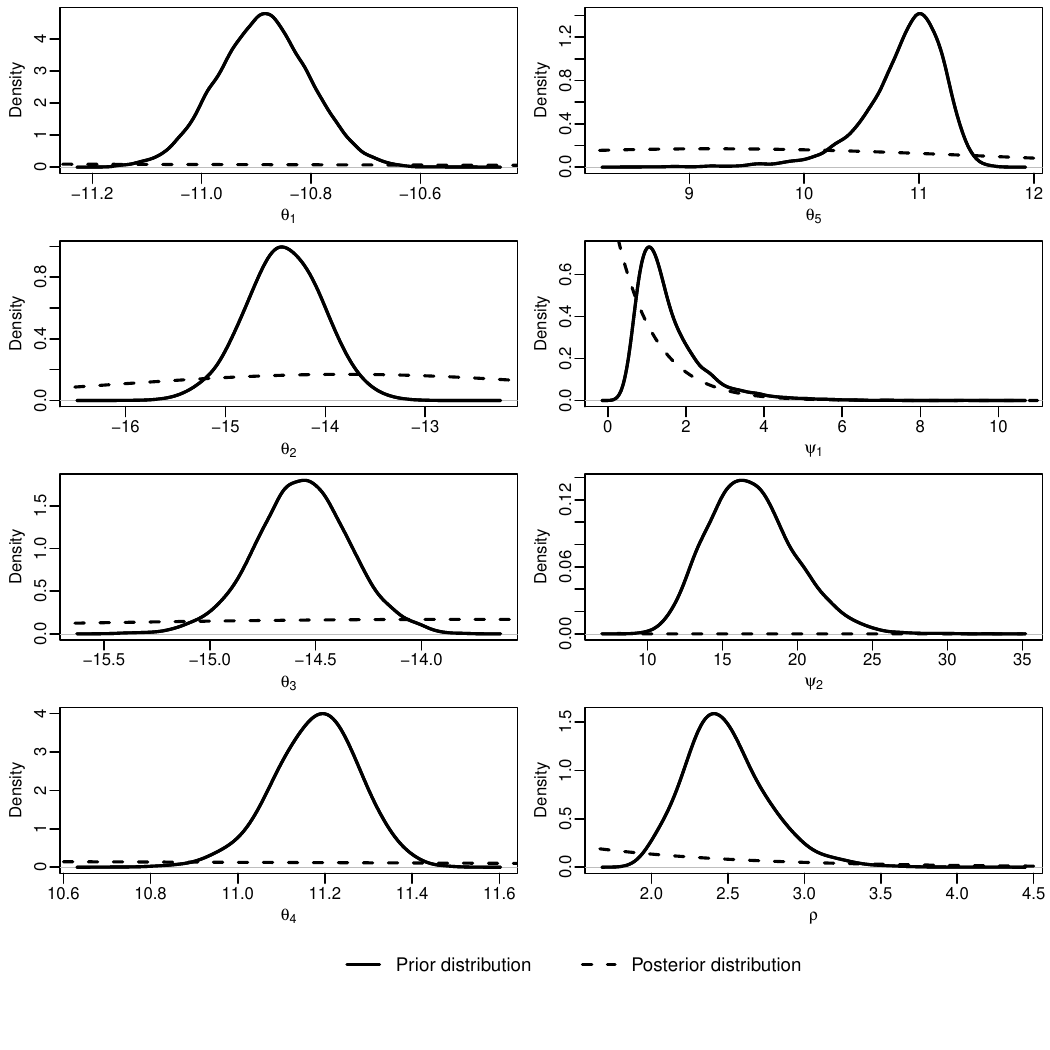} 

}

\caption[Trace plots of the posterior sample (left hand panels) and the estimated posterior and prior densities (right hand panels) for each element of $\boldsymbol{\theta}$]{Trace plots of the posterior sample (left hand panels) and the estimated posterior and prior densities (right hand panels) for each element of $\boldsymbol{\theta}$.}\label{fig:PLOT2}
\end{figure}

\end{knitrout}


Diagnostics for assessing the adequacy of the fitted model \citep[see, for example,][Chapter 6]{G2014} indicated that there were no reasons to believe that the fit was inadequate.

\subsection{Prediction Phase}
We now use the sample generated from the posterior distribution of $\boldsymbol{\delta}$, and, in particular, $\boldsymbol{\theta}$, to construct an emulator $Q(\boldsymbol{\theta}, \mathbf{x}, t)$, as defined in~\eqref{emul}, for the numerical solution to the ODEs. We let $N_1 = 20$, $N_2 = 50$ and $N_3= 50$. It could be noted that this value of $N_1$ is quite small compared to the rule of thumb of \cite{LSW2009} that the sample size should be approximately ten times the number of input dimensions (in this case $p=5$). However, we found that, due to the concentration of the posterior, this value was sufficiently large to produce an adequate emulator and avoid numerical instability in the inversion of $\mathbf{P}_1$. To assess the adequacy of the emulator, we created a test meta-design of the same size (as above, a space filling sub-sample with a candidate list excluding the points from the original meta-design) and implemented the diagnostic methods of \cite{OW2016}. In particular, the root mean squared errors between the numerical solution and the predictive mean were $3.4 \times 10^{-2}$, $0.19$ and $0.40$ for each of the $K=3$ dimensions of $\boldsymbol{\mu}(\boldsymbol{\theta};\mathbf{x},t)$, respectively. The overall coverage of the 95\% predictive intervals, as assessed on the test design, was 89.0\%.

\begin{table}[ht]
\caption{Optimum values of the controllable factors, $\mathbf{x}$ and $t$, as found by ACE.}
\label{optimum}
\begin{center}
\begin{tabular}{lllll} \hline
Name & Symbol & & Value & Unit \\ \hline
Initial amount of A & $A_0$ & $x_1$ & 30.52 & mols\\
Initial amount of D & $D_0$ & $x_2$ & 91.51 & mols\\
Initial amount of E & $E_0$ & $x_3$ & 26.47 & mols\\
Temperature & $\lambda$ & $x_4$ & 313.15 & K\\
Volume & $V$ & $x_5$ & 31.28 & litres \\ 
Time & t & & 199.10 & s \\ \hline 
\end{tabular}
\end{center}
\end{table}

Now we can use emulator $Q(\boldsymbol{\theta}, \mathbf{x}, t)$ to approximate the probability, $\mathrm{P}(\mathbf{y}_0 \in \mathcal{Y}|\mathbf{y}_S)$, of satisfying the specification limits for given $\mathbf{x}$ and $t$. We need to maximize this probability over the space $\mathcal{S} = \mathcal{X} \times \mathcal{T}$. To do this we use the approximate coordinate exchange (ACE; \citealt{OW2015}) algorithm. ACE was originally developed for finding Bayesian optimal experimental designs where the objective function is maximised over a design space. In these situations, the objective function is usually analytically intractable. ACE uses a cyclic ascent algorithm \citep[e.g.][Chapter 7]{lange2013} to maximise the objective function via a univariate Gaussian process emulator based on evaluations of a Monte Carlo approximation to the objective function. We use the implementation of ACE given by the \texttt{R} package \texttt{acebayes} \citep{ace}.  We use 100 random starts of ACE, where the initial value of $(\mathbf{x},t)$ is uniformly generated in $\mathcal{S} \times \mathcal{T}$. The optimum values of $\mathbf{x}$ and $t$, as found by ACE, are shown in Table~\ref{optimum}. The approximate maximum value of $\mathrm{P}(\mathbf{y}_0 \in \mathcal{Y}|\mathbf{y}_S)$ attained was 0.77 (subject to Monte Carlo error). We investigated the sensitivity of the above approach to the choice of prior distributions for $\rho$, $\boldsymbol{\psi}$, $\boldsymbol{\Omega}$ and $\boldsymbol{\Sigma}$. We did this by repeating the above analysis to find the optimal settings for $\mathbf{x}$ and $t$ under different prior specifications. We found the results to be broadly robust to choice of prior distribution and details of this are provided in Appendix~\ref{APP_d}.

To further explore how $\mathrm{P}(\mathbf{y}_0 \in \mathcal{Y}|\mathbf{y}_S)$ depends on $\mathbf{x}$ and $t$, we fix the initial amount of E, temperature and volume at 26.47 mols, 31.28 litres, and 313.5 K, respectively, i.e. the optimum values as found by ACE. We then vary the initial amounts of A and D, and time over the ranges identified in Table~\ref{limits}. The last row of Figure~\ref{fig:PLOT3} shows $\mathrm{P}(\mathbf{y}_0 \in \mathcal{Y}|\mathbf{y}_S)$ for time against the initial amount of A, with different columns for different initial amounts of D. The first three rows show corresponding plots for the marginal probability of satisfying each of the specification limits (\ref{constraints}) on E, F and H, respectively. The trade-off between the objectives given by the specification limits can be clearly seen. To satisfy the limit $E(t)<3$ mols, $t$ has to be large. This is intuitively obvious since the initial amount of E is 26.47 mols so the process will need to progress for some time before the amount of E has decreased enough to satisfy the limit. The opposite is true for H (initial amount of 0 mols), where the optimum time to satisfy $H(t)<3$ mols is for $t$ to be small. This means that there is a very narrow window of values of $t$ that will satisfy all three constraints, as shown in the last row of Figure~\ref{fig:PLOT3}. Note that the optimum values of $\mathbf{x}$ and $t$ (as shown in Table~\ref{optimum}) lie within this narrow window.

\begin{knitrout}
\definecolor{shadecolor}{rgb}{0.969, 0.969, 0.969}\color{fgcolor}\begin{figure}

{\centering \includegraphics[width=6in]{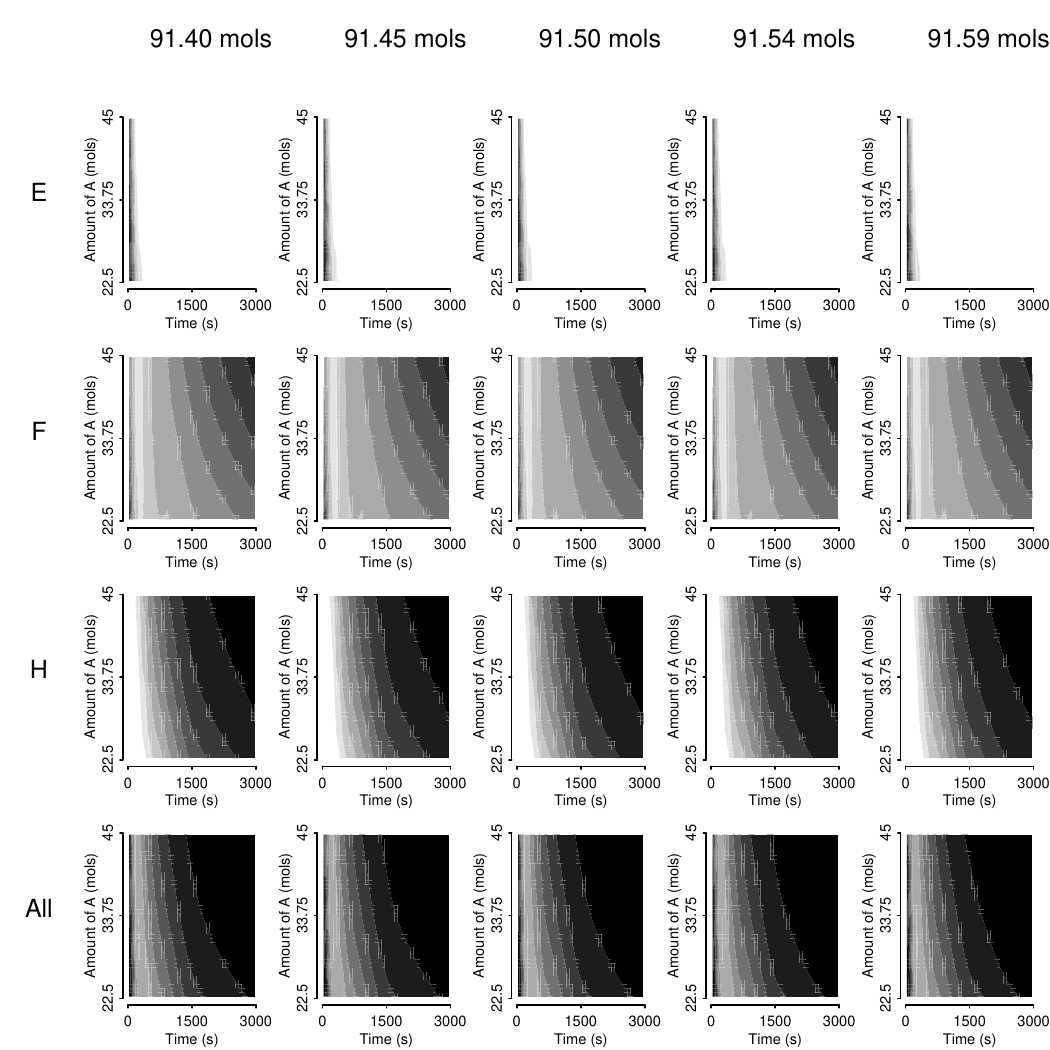} 

}

\caption[The probability of satisfying each of the three marginal constraints and the joint constraints for time (in s]{The probability of satisfying each of the three marginal constraints and the joint constraints for time (in s; $x$-axis) against initial amount of A (in mols; $y$-axis) for five different initial amounts of D. The colours indicate the magnitude of the probability, with black being 0 and white being 1.}\label{fig:PLOT3}
\end{figure}

\end{knitrout}


\section{Discussion}
\label{DISC}
This paper has developed and applied methodology for choosing combinations of values of the controllable factors that produce a high probability of meeting process specification when the process is at least approximating determined by a physical model. Our application was a chemical synthesis process used to produce a pharmaceutical product when specification was determined by a) the amount of the substance constituting the pharmaceutical product being above some level; and b) the amounts of two unwanted by-product substances being below some levels. The relationship between the controllable factors and the amounts of substances of interest is hypothesised to be governed by the analytically intractable solution to a system of ODEs. Responses from a physical experiment are used to refine the knowledge on this relationship and the probability of satisfying the constraints is maximized. Statistical emulators for the numerical solution are constructed to accelerate the model-fitting process and the estimation of the probability of satisfying the constraints. The methodology produces computationally feasible results that take account of different uncertainties involved in the experimental and modelling processes (measurement error, model inadequacy, emulator error). The methods have the potential to be applied to a variety of quality control and pharmaceutical design space problems that involve the numerically expensive approximation to physical models.

The modelling and computational approach taken in this paper can be modified in a variety of different ways to suit the application of interest. For example, we have explicitly taken account of the model discrepancy. If the experimenters believed the model given by the solution to the ODEs was a very close approximation to the true physical process then the model discrepancy could be discarded by setting $\mathbf{D} = \mathbf{0}$ and omitting the steps for sampling from the full conditional distributions of $\boldsymbol{\Sigma}$ and $\boldsymbol{\psi}$. Another example of modification is the introduction of noise into the specification of the controllable factors. In this paper, we assumed that the chemical engineers have complete control over the specification of the controllable factors when we perform the maximization of the probability of satisfying the specification limits given by (\ref{constraints}). In some cases, the chemical engineer would not be able to control these exactly. Following the approach in this paper, it would be straightforward to introduce variability as an intermediate step between specifying the controllable variables and evaluating the Gaussian process emulator. Lastly, we employed the RMLMH algorithm to sample from the full conditional distributions of $\boldsymbol{\theta}$, $\rho$ and $\boldsymbol{\psi}$. It would also be possible to use the methodology in this paper to employ Riemann Manifold Hamiltonian Monte Carlo methodology \citep{GC2011}.

\section*{Acknowledgments}
The authors would like to thank Drs John Peterson, Mohammad Yahyah and Neil Hodnett (GlaxoSmithKline) for providing the original kinetics modelling problem and data set. Additionally, the authors are grateful for the editor, associate editor and two referees who provided valuable feedback on earlier versions of the paper. DCW was support by UK Engineering and Physical Sciences Research Council (EPSRC) Fellowship EP/J018317/1 and KJM was supported by a PhD award from EPSRC and GlaxoSmithKline.

\appendix

\section{Full-conditional distributions for Gibbs sampling}
\label{APP_a}

\subsection{Full conditional distribution for model discrepancy}

For the $r$th chain with temperature $\tau_r$, the full-conditional distribution of $\mathbf{d}^*$ is
$$\mathbf{d}^* | \mathbf{y},\boldsymbol{\theta},\boldsymbol{\Sigma},\boldsymbol{\Omega},\rho,\psi \sim \mathrm{N}\left(\boldsymbol{\mu}_{\mathbf{d}^*},\tau_r \mathbf{V}_{\mathbf{d}^*}\right),$$
where 
\begin{eqnarray*}
\mathbf{V}_{\mathbf{d}^*} & = & \left(\mathbf{H}^{\rm T} \left(\boldsymbol{\Omega}^{-1} \otimes \mathbf{T}^{-1}\right) \mathbf{H} + \left(\boldsymbol{\Sigma}^{-1} \otimes \mathbf{S}^{-1}\right)\right)^{-1},\\
\boldsymbol{\mu}_{\mathbf{d}^*} & = & \mathbf{V}_{\mathbf{d}^*} \left(
\mathbf{H}^{\rm T} \left(\boldsymbol{\Omega}^{-1} \otimes \mathbf{T}^{-1}\right) \mathbf{y} + \left(\boldsymbol{\Sigma}^{-1} \otimes \mathbf{S}^{-1}\right) \mathbf{m}(\boldsymbol{\theta})\right).\end{eqnarray*}

\subsection{Full conditional distributions for covariance matrices}

For the $r$th chain with temperature $\tau_r$, the full-conditional distributions of $\boldsymbol{\Omega}$ and $\boldsymbol{\Sigma}$ are given by
$$\boldsymbol{\Omega}|\mathbf{Y},\mathbf{D}^*,\rho \sim \mathrm{IW} \left(\frac{\nu + n + K + 1}{\tau_r} - K - 1, \frac{1}{\tau_r}\left(\mathbf{I}_K + \left(\mathbf{Y}-\mathbf{GD}^*\right)^{\rm T} \mathbf{T}^{-1}\left(\mathbf{Y}-\mathbf{GD}^*\right)\right)\right),$$
and
$$\boldsymbol{\Sigma}|\mathbf{D}^*,\boldsymbol{\theta},\psi \sim \mathrm{IW} \left(\frac{\nu + m + K + 1}{\tau_r} - K - 1, \frac{1}{\tau_r}\left(\mathbf{I}_k + \left(\mathbf{D}^*-\mathbf{M}(\boldsymbol{\theta})\right)^{\rm T} \mathbf{S}^{-1}\left(\mathbf{D}^*-\mathbf{M}(\boldsymbol{\theta})\right)\right)\right),$$
respectively, where $\mathrm{IW}$ denotes the inverse-Wishart distribution.

\subsection{Full conditional distribution for censored observations}
Let $\mathbf{A}$ be the $n \times n$ permutation matrix which re-orders the elements of $\mathbf{y}$ such that we can write 
$$\mathbf{Ay} = \left(\begin{array}{c} \mathbf{y}_S\\
\mathbf{y}_C \end{array} \right) \sim \mathrm{N}\left( \mathbf{AHd}^*, \mathbf{A}\left(\boldsymbol{\Omega} \otimes \mathbf{T}\right) \mathbf{A}^{\rm T} \right).$$
Define $\mathbf{b} = \mathbf{AHd}^*$ and $\mathbf{L} = \mathbf{A}\left(\boldsymbol{\Omega} \otimes \mathbf{T}\right) \mathbf{A}^{\rm T}$. Let $\mathbf{b} = \left(\mathbf{b}_S,\mathbf{b}_c\right)^{\rm T}$ where $\mathbf{b}_S$ and $\mathbf{b})C$ are the $n_S \times 1$ and $n_c \times 1$ sub-vectors corresponding to $\mathbf{y}_S$ and $\mathbf{y}_C$, respectively. Similarly, partition $\mathbf{L}$ as
$$\mathbf{L} = \left(\begin{array}{cc}
\mathbf{L}_{SS} & \mathbf{L}_{SC} \\
\mathbf{L}_{CS} & \mathbf{L}_{CC} \end{array}\right).$$
The full conditional distribution of $\mathbf{y}_c$ is
$$\mathbf{y}_C \sim \mathrm{N}\left(\mathbf{b}_C + \mathbf{L}_{CS}\mathbf{L}_{SS}^{-1} \left(\mathbf{y}_S - \mathbf{b}_S\right),\tau_r \left(\mathbf{L}_{CC} - \mathbf{L}_{CS}\mathbf{L}_{SS}^{-1}\mathbf{L}_{SC}\right)\right)$$
truncated to the hypercube $\left[-\infty,\log \chi \right]^{n_c}$.

\section{Riemann manifold Langevin Metropolis-Hastings (RMLMH) Algorithm}
\label{APP_b}
Consider generating a sample from the distribution of $\boldsymbol{\delta} \in \mathbb{R}^{p}$, conditional on $\mathbf{y}$, having density $\pi(\boldsymbol{\delta}|\mathbf{y}) \propto \left(\pi(\mathbf{y}|\boldsymbol{\delta}) \pi(\boldsymbol{\delta})\right)^{1/\tau_r}$, for temperature $\tau_r$, using the Metropolis-Hastings algorithm. Let $h(\boldsymbol{\delta}) = \log \pi(\mathbf{y}|\boldsymbol{\delta}) + \log \pi(\boldsymbol{\delta})$ and define
\begin{eqnarray*}
\bigtriangledown h(\boldsymbol{\delta}) & = & \frac{\partial \log \pi(\mathbf{y}|\boldsymbol{\delta})}{\partial \boldsymbol{\delta}} + \frac{\partial \log \pi(\boldsymbol{\delta})}{\partial \boldsymbol{\delta}},\\
\mathbf{G}(\boldsymbol{\delta}) & = & - \mathrm{E}_{\mathbf{y}|\boldsymbol{\delta}} \left[\frac{\partial \log \pi(\mathbf{y}|\boldsymbol{\delta})}{\partial \boldsymbol{\delta}}\frac{\partial \log \pi(\mathbf{y}|\boldsymbol{\delta})}{\partial \boldsymbol{\delta}^{\rm T}}\right] - \frac{\partial^2 \log \pi(\boldsymbol{\delta})}{\partial \boldsymbol{\delta}\partial \boldsymbol{\delta}^{\rm T}}.
\end{eqnarray*}
If $\boldsymbol{\delta}^c$ is the current value of $\boldsymbol{\delta}$, then under the RMLMH algorithm, the proposal distribution is $\mathrm{N}\left(\boldsymbol{\mu}_{\boldsymbol{\delta}},\tau_r\mathbf{V}_{\boldsymbol{\delta}}\right),$ where
\begin{eqnarray*}
\boldsymbol{\mu}_{\boldsymbol{\delta}} & = & \boldsymbol{\delta}^c + \frac{\epsilon^2}{2} \mathbf{G}(\boldsymbol{\delta}^c)^{-1} \bigtriangledown h(\boldsymbol{\delta}^c) - \epsilon^2 \mathbf{n}^{(1)} +  \frac{\epsilon^2}{2} \mathbf{n}^{(2)},\\
\mathbf{V}_{\boldsymbol{\delta}} & = & \epsilon^2 \mathbf{G}(\boldsymbol{\delta}^c)^{-1},
\end{eqnarray*}
with $\mathbf{n}^{(1)}$ and $\mathbf{n}^{(2)}$ having $i$th elements
\begin{eqnarray*}
n^{(1)}_i & = & \sum_{j=1}^p \left\{\mathbf{G}(\boldsymbol{\delta}^c)^{-1} \left.\frac{\partial \mathbf{G}(\boldsymbol{\delta})}{\partial \delta_j} \right\vert_{\boldsymbol{\delta} = \boldsymbol{\delta}^c} \mathbf{G}(\boldsymbol{\delta}^c)^{-1} \right\}_{ij},\\
n^{(2)}_i & = & \sum_{j=1}^p \left\{ \mathbf{G}(\boldsymbol{\delta}^c)^{-1} \right\}_{ij} \mathrm{tr} \left\{ \mathbf{G}(\boldsymbol{\delta}^c)^{-1} \left.\frac{\partial \mathbf{G}(\boldsymbol{\delta})}{\partial \delta_j} \right\vert_{\boldsymbol{\delta} = \boldsymbol{\delta}^c} \right\},
\end{eqnarray*}
respectively.

\section{Components required for Riemann manifold MH Algorithm}
\label{APP_c}

\subsection{Physical parameters}
The log density of the full conditional distribution of $\boldsymbol{\theta}$ is given by
$$h(\boldsymbol{\theta}) \propto - \frac{1}{2}\left(\mathbf{d}^* - \mathbf{m}(\boldsymbol{\theta})\right)^{\rm T} \left(\boldsymbol{\Sigma}^{-1} \otimes \mathbf{S}^{-1}\right)\left(\mathbf{d}^* - \mathbf{m}(\boldsymbol{\theta})\right) - \frac{1}{2}\left(\boldsymbol{\theta} - \boldsymbol{\mu}\right)^{\rm T} \boldsymbol{\Delta}^{-1} \left(\boldsymbol{\theta} - \boldsymbol{\mu}\right).$$
The derivative with respect to $\boldsymbol{\theta}$ is
$$\bigtriangledown h(\boldsymbol{\theta}) = \left(\mathbf{d}^* - \mathbf{m}(\boldsymbol{\theta})\right)^{\rm T} \left(\boldsymbol{\Sigma}^{-1} \otimes \mathbf{S}^{-1}\right)\frac{\partial \mathbf{m}(\boldsymbol{\theta})}{\partial \boldsymbol{\theta}} - \boldsymbol{\Delta}^{-1} \left(\boldsymbol{\theta} - \boldsymbol{\mu}\right).$$
The matrix tensor for $\boldsymbol{\theta}$ is
$$\mathbf{G}(\boldsymbol{\theta}) = \frac{\partial \mathbf{m}(\boldsymbol{\theta})}{\partial \boldsymbol{\theta}^{\rm T}}\left(\boldsymbol{\Sigma}^{-1} \otimes \mathbf{S}^{-1}\right)\frac{\partial \mathbf{m}(\boldsymbol{\theta})}{\partial \boldsymbol{\theta}}+ \boldsymbol{\Delta}^{-1},$$
with derivatives
$$\frac{\partial \mathbf{G}(\boldsymbol{\theta})}{\partial \theta_k} = 2 \frac{\partial \mathbf{m}(\boldsymbol{\theta})}{\partial \boldsymbol{\theta}^{\rm T}}\left(\boldsymbol{\Sigma}^{-1} \otimes \mathbf{S}^{-1}\right)\frac{\partial^2 \mathbf{m}(\boldsymbol{\theta})}{\partial \boldsymbol{\theta} \partial \theta_k},$$
for $k=1,\dots,p$.

\subsection{Correlation parameter for time dependency}
Consider a log transformation of $\rho$, i.e. $a=\log \rho$. The log density of the full conditional distribution of $a$ is given by
\begin{eqnarray*}
h(a) & \propto & -\frac{K}{2}\sum_{i=1}^{I'} \log |\mathbf{T}_i| - \frac{1}{2}\left(\mathbf{y} - \mathbf{Hd}^*\right)^{\rm T} \left(\boldsymbol{\Omega}^{-1} \otimes \mathrm{diag}_{i=1,\dots,I'} \left\{ \mathbf{T}_i^{-1} \right\}\right)\left(\mathbf{y} - \mathbf{Hd}^*\right)\\
& & \qquad \mbox{} + a - \exp(a).
\end{eqnarray*}
The derivative with respect to $a$ is 
\begin{eqnarray*}
\frac{\partial h(a)}{\partial a} & = & -\frac{K}{2}\sum_{i=1}^{I'} \mathrm{tr}\left\{ \mathbf{T}_i^{-1} \mathbf{T}_{ia}\right\} \\
& & + \frac{1}{2}\left(\mathbf{y} - \mathbf{Hd}^*\right)^{\rm T} \left(\boldsymbol{\Omega}^{-1}\otimes \mathrm{diag}_{i=1,\dots,I'} \left\{ \mathbf{T}_i^{-1}\mathbf{T}_{ia} \mathbf{T}_i^{-1} \right\}\right)\left(\mathbf{y} - \mathbf{Hd}^*\right)\\
& & \qquad \mbox{} + 1 - \exp (a),
\end{eqnarray*}
where $\mathbf{T}_{ia} = \partial \mathbf{T}_i / \partial a$ with $jl$th element $-\exp(a)(t_{ij} - t_{il})^2 T_{i,jl}$. The tensor matrix for $a$ is
$$\mathbf{G}(a) = \frac{K}{2} \mathrm{tr} \left\{ \mathbf{T}_i^{-1} \mathbf{T}_{ia} \mathbf{T}_i^{-1} \mathbf{T}_{ia} \right\} - \exp(a),$$
with derivative
$$\frac{\partial \mathbf{G}(a)}{\partial a} = K \mathrm{tr} \left\{ \mathbf{T}_i^{-1} \mathbf{T}_{ia} \mathbf{T}_i^{-1}\left(\mathbf{T}_{iaa} - \mathbf{T}_{ia} \mathbf{T}_i^{-1}  \mathbf{T}_{ia}\right)\right\} - \exp(a),$$
where $\mathbf{T}_{iaa} = \partial^2 \mathbf{T}_i / \partial a^2$ with $jl$th element $\exp(a)(t_{ij} - t_{il})^2 \left(\exp(a)(t_{ij} - t_{il})^2 -1 \right) T_{i,jl}$.

\subsection{Correlation parameters for model discrepancy}
Consider a log transformation of each element of $\boldsymbol{\psi}$, i.e. $b_i = \log \psi_i$ for $i=1,2$. The log density of the full conditional distribution of $\mathbf{b} = (b_1,b_2)$ is given by
$$h(\mathbf{b}) \propto - \frac{K}{2}\log|\mathbf{S}| - \frac{1}{2} \left(\mathbf{d}^* - \mathbf{m}(\boldsymbol{\theta})\right)^{\rm T} \left(\boldsymbol{\Sigma} \otimes \mathbf{S}^{-1}\right)\left(\mathbf{d}^* - \mathbf{m}(\boldsymbol{\theta})\right) + \sum_{i=1}^2 b_i - \exp(b_i).$$
The derivative with respect to $b_i$ is
$$\frac{\partial h(\mathbf{b})}{\partial b_i} = - \frac{K}{2}\mathrm{tr}\left\{ \mathbf{S}^{-1}\mathbf{S}_i\right\} + \frac{1}{2}\left(\mathbf{d}^* - \mathbf{m}(\boldsymbol{\theta})\right)^{\rm T} \left(\boldsymbol{\Sigma} \otimes \mathbf{S}^{-1}\mathbf{S}_i \mathbf{S}^{-1}\right)\left(\mathbf{d}^* - \mathbf{m}(\boldsymbol{\theta})\right) + 1 - \exp(b_i),$$
where $\mathbf{S}_i = \partial \mathbf{S}/\partial b_i$ with $jl$th element given by
$$\left\{
\begin{array}{ll}
- \exp(b_1) \left[\sum_{f=1}^F \left(d_{jf} - d_{lf}\right)^2\right] S_{jl}, & \mbox{if $i=1$},\\
- \exp(b_2) (t_j - t_l)^2 S_{jl}, & \mbox{if $i=2$}.
\end{array}\right.$$
The tensor matrix, $\mathbf{G}(\mathbf{b})$ has $ij$th element
$$G(\mathbf{b})_{ij} = \frac{K}{2} \mathrm{tr}\left\{ \mathbf{S}^{-1}\mathbf{S}_i \mathbf{S}^{-1}\mathbf{S}_j\right\} - I(i=j)\exp(b_i).$$
The derivatives of $\mathbf{G}(\mathbf{b})$ are
\begin{eqnarray*}
\frac{\partial G(\mathbf{b})_{ij}}{\partial b_k} & = & \frac{K}{2} \mathrm{tr}\left\{ \mathbf{S}^{-1} \mathbf{S}_i \mathbf{S}^{-1} \mathbf{S}_{jk} - \mathbf{S}^{-1}\mathbf{S}_i\mathbf{S}^{-1}\mathbf{S}_k\mathbf{S}^{-1}\mathbf{S}_j + \mathbf{S}^{-1} \mathbf{S}_j \mathbf{S}^{-1} \mathbf{S}_{ik} - \mathbf{S}^{-1}\mathbf{S}_i\mathbf{S}^{-1}\mathbf{S}_j\mathbf{S}^{-1}\mathbf{S}_k\right\}\\
& & \qquad \mbox{} - I(i=j=k)\exp(b_i),
\end{eqnarray*}
where $\mathbf{S}_{ik} = \partial^2 \mathbf{S}/\partial b_i \partial b_k$ with $jl$th element given by 
$$\left\{
\begin{array}{ll}
\exp(b_1)\left[\sum_{f=1}^F \left(d_{jf} - d_{lf}\right)^2\right]\left(\exp(b_1)\left[\sum_{f=1}^F \left(d_{jf} - d_{lf}\right)^2\right]-1\right) S_{jl}, & \mbox{if $i=1$ and $k=1$},\\
\exp(b_2)(t_j - t_l)^2\left(\exp(b_2)(t_j - t_l)^2-1\right) S_{jl}, & \mbox{if $i=2$ and $k=2$},\\
\exp(b_1+b_2)\left[\sum_{f=1}^F \left(d_{jf} - d_{lf}\right)^2\right](t_j - t_l)^2 S_{jl}, & \mbox{if $i=1$ and $k=2$}.
\end{array}\right.$$

\section{Prior sensitivity analysis}
\label{APP_d}

Suppose 
$$\begin{array}{lcllcl}
\rho & \sim & \mathrm{Gamma}(\alpha,\beta) & \boldsymbol{\Sigma} & \sim & \mathrm{IW}(\nu, \mathbf{I}_3)\\
\psi_i & \sim & \mathrm{Gamma}(\alpha,\beta) & \boldsymbol{\Omega} & \sim & \mathrm{IW}(\nu, \mathbf{I}_3)
\end{array}$$
for $i=1,2$, where the Gamma distribution is parameterised such that the expectation and variance are $\alpha/\beta$ and $\alpha/\beta^2$, respectively, and the inverse-Wishart such that the mode is $\mathbf{I}_3/\nu$. Let the prior distribution for $\rho$, $\boldsymbol{\psi}$, $\boldsymbol{\Sigma}$ and $\boldsymbol{\Omega}$ outlined in Section~\ref{sec:model} be denoted as Prior 0 where $\alpha = \beta = 1$ and $\nu = 4$. We consider three different prior specifications (denoted Prior 1, 2 and 3) given by different combinations of $\alpha$, $\beta$ and $\nu$, corresponding to a sequence of increasingly diffuse prior distributions. For each prior, the analysis described in Section~\ref{RES} was undertaken and the optimum values of $\mathbf{x}$ and $t$ were found by ACE. Table~\ref{tab:sens} shows the values of $\alpha$, $\beta$ and $\nu$ for Priors 1, 2 and 3 as well as the optimum values of $\mathbf{x}$ and $t$. Also included is the probability of satisfying the specification limits given by (\ref{constraints}) under each prior distribution for the optimum values also found under each prior distribution. There is some variability with respect to initial amounts of A and D and, particularly, with respect to time. However the probability of satisfying the specification limits remain broadly insensitive to these changes. We conclude that the results are robust to choice of prior distribution.

\begin{table}[ht]
\caption{For Priors 0, 1 , 2 and 3: a) the values of $\alpha$, $\beta$ and $\nu$ specifying the prior distributions for $\boldsymbol{\psi}$, $\boldsymbol{\Sigma}$ and $\boldsymbol{\Omega}$; b) the optimum values of $\mathbf{x}$ and $t$ for maximizing the probability of satisfying the specification limits; and c) the probability of satisfying the specification limits given by (\ref{constraints}) under each prior distribution for the optimum values also found under each prior distribution.}
\label{tab:sens}
\begin{center}
\begin{tabular}{lllll} \hline
Prior & 0 & 1 & 2 & 3 \\ \hline
$\alpha$ & 1 & 0.500 & 0.250 & 0.125 \\
$\beta$ & 1 & 0.500 & 0.250 & 0.125 \\
 $\nu$ & 4 & 2 & 1 & 0.5 \\ \hline
 \multicolumn{5}{c}{Optimum values for controllable factors and time} \\ \hline
 Initial amount of A & 30.52 & 22.50 & 39.58 & 37.75 \\
Initial amount of D & 91.51 & 91.49 & 91.59 & 91.59 \\
Initial amount of E & 26.47 & 26.47 & 26.47 & 26.47 \\
Temperature & 313.15 & 313.15 & 313.15 & 313.15 \\
Volume & 31.28  & 31.28 & 31.28 & 31.28 \\
Time & 199.10 & 453.36 & 169.59 & 169.59 \\ \hline
\multicolumn{5}{c}{Probability of satisfying specification limits} \\ \hline
Prior 0 & 0.77 & 0.75 & 0.78 & 0.73 \\
Prior 1 & 0.74 & 0.75 & 0.75 & 0.73 \\
Prior 2 & 0.76 & 0.74 & 0.78 & 0.73 \\
Prior 3 & 0.76 & 0.74 & 0.77 & 0.76 \\ \hline 
\end{tabular}
\end{center}
\end{table}


\bibliographystyle{model2-names}
\bibliography{mybib}

\end{document}